\documentstyle[12pt,fleqn]{article}
\begin{document}
\baselineskip=7.5mm

\begin{center}
{\large\bf Gamma-Ray Burst Afterglows from Realistic Fireballs}

\vspace{5.0mm}
Z. G. Dai, Y. F. Huang, and T. Lu

{\em Department of Astronomy, Nanjing University, Nanjing 210093, China}
\end{center}

\vspace{3mm}

\begin{center}
ABSTRACT
\end{center}

A GRB afterglow has been commonly thought to be due to continuous
deceleration of a postburst fireball. Many analytical models
have made simplifications for deceleration dynamics of
the fireball and its radiation property, although they are successful
at explaining the overall features of the observed afterglows.
We here propose a model for a GRB afterglow in which the evolution
of a postburst fireball is in an intermediate
case between the adiabatic and highly radiative expansion. In
our model, the afterglow is both due to the contribution of the adiabatic
electrons behind the external blastwave of the fireball and due to
the contribution of the radiative electrons. In addition,
this model can describe evolution of the fireball from the extremely
relativistic phase to the non-relativistic phase. Our calculations show 
that the fireball will go to the adiabatic expansion phase after
about a day if the accelerated electrons are assumed to occupy the total
internal energy. In all cases considered, the fireball will go to the
mildly relativistic phase about $10^4$ seconds later, and to the
non-relativistic phase after several days. These results imply that
the relativistic adiabatic model cannot describe the deceleration
dynamics of the several-days-later fireball. The
comparison of the calculated light curves with the observed results
at late times may imply the presence of impulsive events or energy
injection with much longer durations.

\noindent
{\em Subject headings:} gamma-ray : bursts --- radiation
                        mechanisms : nonthermal

\newpage

\begin{center}
1. INTRODUCTION
\end{center}

Even though the energy source for gamma-ray bursts (GRBs) has
remained unknown, the popular theoretical explanation for their radiative
properties has been commonly thought to be the fireball+shock wave
model, in which a GRB results from the dissipation of the kinetic
energy of a relativistically expanding fireball. This dissipation
can be either (most likely) due to internal shocks produced
during the collision between the shells with different Lorentz factors
in the fireball (Rees \& M\'esz\'aros 1994; Paczy\'nski \& Xu 1994; Sari \&
Piran 1997), or due to external shocks (a forward blastwave and a reverse
shock) formed by the fireball colliding with the surrounding medium
(Rees \& M\'esz\'aros 1992; M\'esz\'aros \& Rees 1993; Katz 1994; Sari,
Narayan, \& Piran 1996). After the main GRB, the fireball will
continuously decelerate due to more and more swept-up medium matter
and therefore will produce delayed emission at longer wavelengths,
an afterglow, as predicted in advance of the observations
(Paczy\'nski \& Rhoads 1993; Katz 1994; M\'esz\'aros \& Rees 1997;
Vietri 1997a).

Afterglows from GRBs have been observed from
a number of objects at X-ray, optical, and in one case also at radio
wavelengths. The published analytical models are successful at explaining
the major features of the light curves (M\'esz\'aros \& Rees 1997;
Wijers, Rees, \& M\'esz\'aros 1997; Reichart 1997; Waxman 1997a,b;
Tavani 1997; Vietri 1997a,b; Katz \& Piran 1997; Dai \& Lu 1998a).
However, such models cannot provide a {\em detailed} description
for the evolution of a postburst fireball and the light curve of an
afterglow because they have made simplifications in three aspects.
First, all of these models have assumed that the postburst fireball is
extremely relativistic. Second, Wijers et al. (1997), Waxman (1997a,b),
and Reichart (1997) considered the adiabatic expansion of the fireball
in uniform interstellar medium. This is a reasonable assumption
if the timescale for cooling of the accelerated electrons behind
the blastwave is much longer than the expansion timescale of the fireball
or the electrons carry a small fraction of the internal energy.
This simple model has given a scaling relation between the fireball's
Lorentz factor ($\gamma$) and the blastwave's radius ($R$): $\gamma\propto
R^{-3/2}$. Dai \& Lu (1998a) further discussed the effect of radiative
corrections and nonuniformity of the medium on GRB afterglows.
On the other hand, Vietri (1997a,b) postulated that the postburst
fireballs are highly radiative. This requires that the accelerated
electrons behind the blastwave occupy all the internal energy, and
that they always cool much more rapidly than the fireball expands.
Such a model has given another scaling relation: $\gamma\propto R^{-3}$.
Third, it is usually assumed that the distribution of the electrons
behind a shock is a power law. Since electrons with different
Lorentz factors should have different efficiencies for synchrotron
radiation in the same magnetic field, the spectrum radiated from
higher-energy electrons for this distribution is steeper than that
from lower-energy electrons (Sari, Piran, \& Narayan 1998).
The adiabatic expansion model (Wijers et al. 1997; Waxman 1997a,b;
Reichart 1997; Dai \& Lu 1998a) has assumed all of the accelerated electrons
behind the blastwave to be adiabatic, while in the radiative expansion model
(Vietri 1997b) these electrons has been thought to be highly radiative.

In fact, the fireball first expands relativistically, and will
eventually go into the non-ralativistic phase (the Sedov phase)
after a long time. Furthermore, the actual expansion of the fireball
is likely in an intermediate case between the adiabatic and highly radiative
expansion. Finally, an afterglow may be contributed by both
the adiabatic electrons and radiative electrons behind the blastwave.
In this paper we would propose a model which addresses the above
issues. Huang et al. (1998) first studied numerically the evolution
of an adiabatic fireball from the ultrarelativistic
expansion phase to mildly relativistic expansion phase.
The present work is in fact a significant development of the study of
Huang et al. (1998) through considering the above three issues.
This paper is organized as follows: in section 2 we
calculate numerically the dynamical evolution of a postburst fireball
from the ultrarelativistic expansion phase to Sedov phase. In section 3
we formulate synchrotron radiation from the accelerated electrons
behind the blastwave and compare our results with observations, and
in the final section we give a brief discussion.

\begin{center}
2. HYDRODYNAMICS OF POSTBURST FIREBALLS
\end{center}

We assume that a fireball with an amount of energy $E$ comparable to
that observed in gamma rays, $E\sim 10^{51}$--$10^{52}$\,ergs,
and with the mass of the contaminating baryons, $M_0$, is
produced, and after an initial acceleration its Lorentz factor is
$\eta=E/(M_0c^2)$. Subsequently, at the radius $R_0$, the expansion of
the fireball starts to be significantly influenced by the swept-up medium
and two external shocks may form: a forward blastwave and a reverse shock
(Rees \& M\'esz\'aros 1992). As usual, $R_0$ is supposed to be
\begin{equation}
R_0=\left(\frac{3E}{4\pi n m_pc^2\eta^2}\right)^{1/3}
=10^{16}E_{51}^{1/3}n_0^{-1/3}\eta_{300}^{-2/3}\,\,\,{\rm cm}\,,
\end{equation}
where $E_{51}=E/10^{51}\,{\rm ergs}$, $\eta_{300}=\eta/300$, and $n_0$
is the electron number density of the medium in units of 1\,cm$^{-3}$
($n_0=n/1\,{\rm cm}^{-3}$). Following the main GRB event, which may be
produced by nonthermal processes such as synchrotron or
possibly inverse Compton emission, the blastwave continues to
sweep up the medium.

According to Blandford \& McKee (1976), the electron number density ($n'$)
and energy density ($e'$) of the shocked medium in the frame comoving with  
the fireball and the Lorentz factor of the blastwave ($\Gamma$) can be 
written as
\begin{equation}
n'=\frac{\hat{\gamma}\gamma+1}{\hat{\gamma}-1}n\,,
\end{equation}
\begin{equation}
e'=\frac{\hat{\gamma}\gamma+1}{\hat{\gamma}-1}(\gamma-1)nm_pc^2\,,
\end{equation}
\begin{equation}
\Gamma=\left\{\frac{(\gamma+1)[\hat{\gamma}(\gamma-1)+1]^2}
       {\hat{\gamma}(2-\hat{\gamma})(\gamma-1)+2}\right\}^{1/2}\,,
\end{equation}
where $\hat{\gamma}$ is the adiabatic index of the shocked medium, which is
generally between 4/3 and 5/3. One expects that equations (2)--(4) are
appropriate for both relativistic and non-relativistic blastwaves. From
the definition of $\hat{\gamma}$ (Blandford \& McKee 1976), we have derived
an expression: $\hat{\gamma}\approx (4\gamma+1)/(3\gamma)$. It can be
seen from this approximation that $\hat{\gamma}\approx 4/3$ for an extremely
relativistic blastwave and $\hat{\gamma}\approx 5/3$ for the Sedov shock.

We further assume that the magnetic energy density in the comoving frame 
is a fraction $\xi_B^2$ of the total thermal energy density, viz., 
$B^\prime=\xi_B(8\pi e^\prime)^{1/2}$, and that the accelerated electrons 
behind the blastwave carry a fraction $\xi_e$ of the energy. This implies
that the minimum Lorentz factor of the random motion of electrons
in the comoving frame is $\gamma_{min}=\xi_e(\gamma-1) m_p/m_e+1$. We here 
consider only synchrotron emission from these electrons, and neglect 
the contribution of inverse Compton emission because the latter emission is
not of importance particularly at late times of the evolution (Waxman 1997a; 
Dai \& Lu 1998a). The energy of a typical accelerated electron behind the
blastwave is lost both through synchrotron radiation and through expansion of
the fireball, and thus the radiative efficiency of this {\em electron} is
given by $t^{\prime -1}_{syn}/(t^{\prime -1}_{syn}+t^{\prime -1}_{ex})$,
where $t^\prime_{syn}$ is the synchrotron cooling time, $t^\prime_{syn}
=6\pi m_ec/(\sigma_TB^{\prime 2}\gamma_{min})$, 
and $t^\prime_{ex}$ is the comoving-frame expansion time, 
$t^\prime_{ex}=R/(\gamma c)$ (Dai \& Lu 1998a). Here $R$ is the radius of 
the blastwave. Since all of the accelerated electrons behind the blastwave
carry only a fraction $\xi_e$ of the internal energy, the radiative
efficiency of the {\em fireball} can be given by
\begin{equation}
f=\xi_e \frac{t^{\prime -1}_{syn}}{t^{\prime -1}_{syn}+t^{\prime -1}_{ex}}\,.
\end{equation}
For the adiabatic expansion, $\xi_e\ll 1$ or $t^{\prime}_{syn}\gg
t^\prime_{ex}$, so $f\approx 0$; but for the highly radiative expansion,
$\xi_e\approx 1$ and $t^{\prime}_{syn}\ll t^\prime_{ex}$ lead to
$f\approx 1$. One expects that in the intermediate case the radiative
efficiency of the fireball ($f$) is between 0 and 1.

In the absence of radiation and expansion losses, according to equations (2)
and (3), the kinetic energy per baryon of the shocked medium in the comoving
frame is $(\gamma-1)m_pc^2$. But, in the presence of these losses, such a
kinetic energy becomes $(\gamma-1)(1-f)m_pc^2$. Thus, the total kinetic
energy of the shocked medium in the burster's rest frame is
$\{\gamma [(\gamma-1)(1-f)+1]-1 \}Mc^2$, where $M$ is the mass of
the swept-up medium, $M=(4\pi/3)R^3nm_p$. We therefore obtain the total
kinetic energy of the fireball
\begin{equation}
E_k=\{\gamma [(\gamma-1)(1-f)+1]-1 \}Mc^2 +(\gamma-1)M_0c^2\,,
\end{equation}
where the second term is the kinetic energy of the contaminating baryons.
If this term is neglected and $f\approx 0$, then $E_k\approx \gamma^2Mc^2$
for extremely relativistic expansion. This expression is just
the starting point of many works (e.g., M\'esz\'aros \& Rees 1997;
Wijers et al. 1997; Waxman 1997a,b; Dai \& Lu 1998a). If $f\approx 1$,
then $E_k \approx (\gamma-1)(M+M_0)c^2$, which turns out to be the case
discussed by Vietri (1997a,b) and Katz \& Piran (1997).
It should be pointed out that at late times the kinetic energy of the
fireball is $E_k\approx (\gamma^2-1)Mc^2$, which is a factor of 1.4 larger
than the Sedov result (Blandford \& McKee 1976). Hence, equation (6) can
also describe well the non-relativistic evolution of the fireball.
Due to synchrotron radiation, the rate for the kinetic energy loss
is given by (Blandford \& McKee 1976)
\begin{equation}
\frac{dE_k}{dt_b}=-4\pi R^2(\beta c)\gamma (\gamma-1) nm_pc^2 f\,,
\end{equation}
where $t_b$ is the burster-rest-frame time and $\beta=(1-\gamma^{-2})^{1/2}$.

In order to study dynamical evolution of the fireball, we should
add two differential equations (Huang et al. 1998):
\begin{equation}
\frac{dR}{dt_b}=(1-\Gamma^{-2})^{1/2}c\,,
\end{equation}
\begin{equation}
\frac{dt_b}{dt}=\gamma (\gamma+\sqrt{\gamma^2-1})\,,
\end{equation}
where $t$ is the observer-frame time.

Equations (3)--(9) present a perfect description for the dynamical evolution
of the postburst fireball. In order to solve these equations, we must
determine the initial conditions. Assuming that $\gamma_0$ and $E_{k0}$
are the initial values of the Lorentz factor and kinetic energy of
the fireball respectively, we require
\begin{equation}
E_{k0}=\frac{E}{2}=\{\gamma_0 [(\gamma_0-1)(1-f_0)+1]-1\}(M_0c^2/\eta)+
       (\gamma_0-1)M_0c^2\,,
\end{equation}
where $f_0$ is the initial value of the radiative efficiency.
Let's define an index $\alpha$ through the following expression:
\begin{equation}
\alpha\equiv -\frac{d\ln \gamma}{d\ln R}\,.
\end{equation}
We take $E=10^{51}\,$ergs, $n=1\,{\rm cm}^{-3}$, and $M_0=10^{-6}M_\odot$
and $2\times 10^{-6}M_\odot$, and our numerical results are shown
in Figures 1--7. Figure 1 illustrates evolution of the fireball's kinetic
energy, and Figures 2 and 3 give evolution of $\gamma (t)$ and $R(t)$.
Figure 2 shows that the fireballs will go to the
mildly relativistic phase about $10^4$ seconds later, and to the
non-relativistic phase after several days, implying that
the relativistic adiabatic model cannot describe the deceleration
dynamics of a several-days-later fireball. We have compared our numerical
result with the analytical solution for the Sedov phase, and found that
they are in good agreement at the non-relativistic phase. This comparison
is shown in Figure 4. In Figures 5 and 6 the time-dependence of $f$ and
$\alpha$ is plotted.
It can be seen from Figure 5 that in the case of $\xi_e=1$ (the solid,
dotted and dashed lines) $f$ is first kept to be a constant ($\approx 1$),
subsequently declines quickly, and finally tends to zero after
$10^6$ seconds, showing that the fireball first expands radiatively,
soon later goes into the intermediate expansion phase, and finally becomes
a non-relativistic shock. In this case, $\alpha$ first increases up to a peak
near 3 due to the influence of the contaminating baryons, and then
decreases to 1.5 at about 1 day. After this, $\alpha$ decreases
to zero because the Lorentz factor becomes one during the non-relativistic
phase. Figures 7 gives evolution of the mass of the swept-up medium.

\begin{center}
3. X-RAY AND OPTICAL RADIATION
\end{center}

In the absence of radiation loss, the distribution of the accelerated
electrons behind the blastwave is usually assumed to be a power-law
function of electron energy:
\begin{equation}
\frac{dN_e}{d\gamma_e}\propto \gamma_e^{-p},
\,\,\,\,\,\,{\rm for}\,\,\gamma_{min}\le \gamma_e\le\gamma_{max}\,,
\end{equation}
where $\gamma_{max}$ is the maximum Lorentz factor, $\gamma_{max}=
10^8(B^{\prime}/1{\rm G})^{-1/2}$ (Dai \& Lu 1998a), and $p$ is the index
between 2 and 3. However, radiation loss may modify such a simple
distribution. In a magnetic field, electrons with different Lorentz
factors have different efficiencies for synchrotron radiation. As defined
by Sari et al. (1998), the critical electron Lorentz factor, $\gamma_c$,
above which synchrotron radiation is significant, is written as
\begin{equation}
\gamma_c=\frac{3m_e}{16\xi_B^2\sigma_Tm_pc}\,\frac{1}{t\gamma^3n}\,.
\end{equation}
Electrons with Lorentz factors below $\gamma_c$ are referred to as
adiabatic ones, and electrons above $\gamma_c$ as radiative ones.
In the presence of steady injection of electrons accelerated by the shock,
the distribution of radiative electrons becomes another power-law
function with an index of $p+1$ (Rybicki \& Lightman 1979), but the
distribution of adiabatic electrons is unchanged. Thus, the actual
distribution can be given in three cases: (i) For $\gamma_c\le \gamma_{min}$,
\begin{equation}
\frac{dN_e}{d\gamma_e}=C_1\gamma_e^{-(p+1)}\,, \,\,\,\,\,\,
C_1=\frac{p}{\gamma_{min}^{-p}-\gamma_{max}^{-p}}N_{tot}\,, \,\,\,\,\,\,
(\gamma_{min}\le\gamma_e\le\gamma_{max})\,,
\end{equation}
where $N_{tot}$ is the total electron number of the shocked medium
($N_{tot}=M/m_p$).
(ii) For $\gamma_{min}< \gamma_c < \gamma_{max}$,
\begin{equation}
  \frac{dN_e}{d\gamma_e} = \left \{
   \begin{array}{ll}
 C_2\gamma_e^{-p}\,, \,\,\,\,\,\, & \gamma_{min}<\gamma_e\le\gamma_c \\
 C_3\gamma_e^{-(p+1)}\,, \,\,\,\,\,\, & \gamma_c<\gamma_e<\gamma_{max}\,,
   \end{array}
   \right. 
\end{equation}
where 
\begin{equation}
 C_2=C_3/\gamma_c\,,
\end{equation}
\begin{equation}
 C_3=\left[\frac{\gamma_{min}^{1-p}-\gamma_c^{1-p}}{\gamma_c(p-1)}+
     \frac{\gamma_c^{-p}-\gamma_{max}^{-p}}{p} \right]^{-1}N_{tot} \,.
\end{equation}
(iii) If $\gamma_c\ge \gamma_{max}$, then
\begin{equation}
\frac{dN_e}{d\gamma_e}=C_4\gamma_e^{-p}\,, \,\,\,\,\,\,
C_4=\frac{p-1}{\gamma_{min}^{1-p}-\gamma_{max}^{1-p}}N_{tot}
\,, \,\,\,\,\,\, (\gamma_{min}\le\gamma_e\le\gamma_{max})\,.
\end{equation}

After having the modified electron distribution functions, we can calculate
the radiation flux. The power for synchrotron radiation from all the
accelerated electrons of the shocked medium in the comoving
frame is given by
\begin{equation}
     j(\nu^\prime)\equiv
     \frac {dP^\prime}{d\nu^\prime} = \frac {\sqrt{3} e^3B'}
     {m_e c^2} \int_{\gamma_{min}}^{\gamma_{max}} 
     \frac{dN_e}{d\gamma_e} F({\nu^\prime}/{\nu_c^\prime}) d\gamma_e\,,
\end{equation}
where
\begin{equation}
F(x) = x \int_{x}^{+\infty} K_{5/3}(t) dt,
\end{equation}
and 
\begin{equation}
\nu_c^\prime = \frac {\gamma_{e}^2 e B'}{2 \pi m_e c}
\end{equation}
with $K_{5/3}(t)$ being the Bessel function. The observer-frame flux
density should be
\begin{equation}
S_\nu=\gamma(1+\beta)^3 j\left(\frac{\nu}{\gamma(1+\beta)}\right)
      \frac{1}{4\pi D^2}\,,
\end{equation}
where $D$ is the source distance to the observer. In writing
equation (22), we have assumed that the emitting equal-time surface of
the source is an ellipsoid. However, the actual fireball always
decelerates due to more and more swept-up medium. It has been found that
due to the deceleration the emitting surface becomes a distorted
ellipsoid (egg-like shape) (Panaitescu \& M\'esz\'aros 1998; Sari 1998),
which slightly influences the observer-frame flux density.
Huang et al. (1998) considered this effect by introducing a factor near
five. Here we would neglect this effect. The flux observed by an X-ray
detector is an integral of $S_\nu$:
\begin{equation}
F_{ob}(t) = \int_{\nu_l}^{\nu_u}S_\nu d\nu\,,
\end{equation}
where $\nu_u$ and $\nu_l$ are the upper and lower frequency limits of
the detector. 

In the previous section, the dynamical evolution of a postburst fireball 
has been evaluated numerically. Now we continue to calculate the 
afterglows at X-ray and optical wavelengths. While some of the 
parameters are fixedly taken ($E = 10^{51}$ ergs, $n = 1$ cm$^{-3}$, 
$p = 2.1$), we change the others such as $M_0$, $\xi_B^2$ and 
$\xi_e$ in reasonable ranges so as to investigate their influence. 
The results are plotted and compared with observations in Figures 8 and 9. 
In these figures, the solid lines correspond 
to $M_0 = 2 \times 10^{-6}$ $M_{\odot}$, $\xi_B^2 = 0.01$, $\xi_e = 1$; 
the dotted lines correspond  
to $M_0 = 1 \times 10^{-6}$ $M_{\odot}$, $\xi_B^2 = 0.01$, $\xi_e = 1$; 
the dashed lines correspond 
to $M_0 = 2 \times 10^{-6}$ $M_{\odot}$, $\xi_B^2 = 0.1$, $\xi_e = 1$; 
and the dash-dotted lines correspond  
to $M_0 = 2 \times 10^{-6}$ $M_{\odot}$, $\xi_B^2 = 0.01$, $\xi_e = 0.5$.  
Figure 8a is the R band afterglow from GRB 970228 and Figure 8b 
illustrates the optical afterglow from GRB 970508. 
Although we have assumed different distances in Figures 8a and 8b, it is not
necessary to imply that GRB 970228 lies farther from us than
GRB 970508 does, since other intrinsic parameters such
as $E$, $n$ may be different. X-ray afterglows or initial X-ray bursts 
were detected for nine
GRBs. Table 1 lists all the flux data available in the literature. In 
Figure 9, X-ray afterglows from these GRBs are plotted together. 
Please note that since different detectors work in different bands, here 
we have converted the flux data into $0.1 - 10$ keV band linearly, errors 
to a factor of two are thus possible. 

We see that the present model generally fits the observations and small
changes of parameters do not alter the overall properties of afterglows. 
However, a problem appears at later times ($t \geq 10^7$ s) after the 
fireball becomes non-relativistic. The model predicts a sharper decline 
while the observed fluxes are obviously higher. In fact, the optical 
afterglow from GRB 970228 was observed to follow approximately a 
power-law decay for at least $\sim 190$ days (Fruchter, Bergeron, \& 
Pian 1997) and the afterglow from GRB 970508 decayed even more slowly 
after about 80 days, implying the presence of a constant component 
(Pedersen et al. 1998; Galama et al. 1998a). The overall power-law 
decay lasting for several months is usually considered as strong evidence 
for the fireball+blastwave model. Here we would like to stress that 
the blastwave in a simple fireball model will cease to be relativistic 
after $\sim 10^6$ s. At least, it is problematic to assume a blastwave 
with $\gamma \geq 5$ when $t \geq 10^7$ s (also see Huang et al. 
1998). One should be cautious in applying the simple scaling laws 
such as $\gamma \propto t^{-3/8}$, $R \propto t^{1/4}$, and 
$S_{\nu} \propto t^{3(1-p)/4}$ in the adiabatic expansion model at such 
late times. So the long-term optical afterglows have really 
raised a problem to the popular fireball+blastwave model. 
For GRB 970508, the optical flux peaked about two days later, possibly 
associated with an X-ray outburst (Piro et al. 1998). This feature 
could not be explained by a simple fireball+blastwave model, too. 
Our theoretical light curve peaks several hours later at optical 
wavelength and several tens of seconds later in X-rays.

\begin{center}
4. DISCUSSION
\end{center}

In this paper
we have tried to propose a model for a GRB afterglow in which the evolution
of a postburst is in an intermediate case
between the adiabatic and highly radiative expansion. In
this model the afterglow is due to the contributions both of the adiabatic
electrons and of the radiative electrons behind the blastwave. In addition,
our model is valid both for the extremely relativistic phase and for
the Sedov phase. Our calculations show that the postburst fireball will
go to the adiabatic expansion phase after about a day if the accelerated
electrons are assumed to occupy the total internal energy (viz.,
$\xi_e=1$). In all cases considered, the fireball will go to the
mildly relativistic phase about $10^4$ seconds later, and to the
non-relativistic phase after several days. These results imply that
the relativistic adiabatic model ($\gamma\propto t^{-3/8}$ and $R\propto
t^{1/4}$) isn't suitable to describe detailedly the deceleration dynamics
of the several-days-later fireball. What we would like to emphasize is
that one should be cautious in applying the simple scaling laws
such as $\gamma \propto t^{-3/8}$, $R \propto t^{1/4}$, and 
$S_{\nu} \propto t^{3(1-p)/4}$ in the adiabatic expansion model at
$t\ge 10^5$ seconds.

In comparing our model with observations, we find that for GRB 970228 the
initial fireball can produce the observed optical afterglow, but the
flux of the fireball after several days is below the observed data,
since the observed afterglow follows approximately
a power-law decay for at least $\sim 190$ days. The optical afterglow
from GRB 970508 first declined slowly, then rised to a peak at about
two days and after this it decayed. But the light curve became flatter
after 80 days than the power-law decay. Comparison of these observational
results with our calculations may imply the presence of impulsive
events or energy injection with much longer durations. One possibility
is that a postburst fireball contains shells with a continuous
distribution of Lorentz factors (Rees \& M\'esz\'aros 1998).
As the external blastwave sweeps up ambient matter and decelerates,
internal shells will eventually catch up with the blastwave
and supply energy into it. A detailed calculation shows that
this model can explain well the afterglow from GRB 970508
(cf. Panaitescu, M\'esz\'aros \& Rees 1998)

Another possibility is that a strongly magnetized millisecond pulsar can
supply its rotational energy into a postburst fireball
(Dai \& Lu 1998b). Many energy-source models of GRBs 
all predicted that as an extremely relativistic fireball is produced,
a strongly magnetized millisecond pulsar is born. It is natural to expect
that magnetic dipole radiation from the pulsar may influence evolution
of the external fireball because the electromagnetic waves are always
absorbed by the shocked medium. Such an effect has been analytically shown
to be able to provide a satisfactary explanation for the flattening
behavior of the light curve of the optical afterglow from GRB 970228 (Dai
\& Lu 1998b). A recent analysis (Dai \& Lu 1998c) further shows that this effect
can also explain well the decline-rise-decline feature of the light curve of
the optical afterglow from GRB 970508 if the index ($p$) of the power-law
distribution of the accelerated electrons behind the blastwave increases
from initial $p=1$ to $p=2.2$ two days later, inferred from the observed
spectrum (Galama et al. 1998a,b).

\vspace{5mm}
This work was supported by the National Natural Science Foundation of China.

\newpage
\baselineskip=4mm

\begin{center}
REFERENCES
\end{center}

\begin{description}
\item Antonelli, L. A., Butler, R. C., Piro, L., Celidonio, G., Coletta, A., 
        Tesseri, A., \& Libero, C. D. 1997, IAU Circ. 6792
\item Blandford, R. D., \& McKee, C. F. 1976, Phys. Fluids, 19, 1130
\item Coletta, A., Gandolfi, G., Smith, M., Piro, L., Cinti, M., Soffitta, P.,
        \& Heise, J. 1997, IAU Circ. 6796
\item Costa, E. et al. 1997a, IAU Circ. 6533
\item Costa, E. et al. 1997b, IAU Circ. 6572
\item Costa, E. et al. 1997c, IAU Circ. 6576
\item Costa, E. et al. 1997d, IAU Circ. 6649
\item Dai, Z. G., \& Lu, T. 1998a, MNRAS, in press
\item Dai, Z. G., \& Lu, T. 1998b, A\&A, 333, L87
\item Dai, Z. G., \& Lu, T. 1998c, Phys. Rev. Lett., submitted
\item Feroci, M. et al. 1997, IAU Circ. 6610
\item Frontera, F. et al. 1997, IAU Circ. 6637
\item Fruchter, A., Bergeron, L., \& Pian, E. 1997, IAU Circ. 6747
\item Galama, T. J. et al. 1998a, ApJ, 497, L13
\item Galama, T. J., Wijers, R. A. M. J., Bremer, M., Groot, P. J.,
         Strom, R. G., Kouvelioto, C., \& van Paradijs, J. 1998b,
         preprint: astro-ph/9804191
\item Garcia, M. R. et al. 1997, preprint: astro-ph/9710346
\item Greiner, J. 1997, IAU Circ. 6742 
\item Greiner, J. 1998, preprint: astro-ph/9802222  
\item Greiner, J., van Paradijs, J., Marshall, F. M., Hurley, K., 
        Robinson, C. R., \& Siebert, J. 1997a, IAU Circ. 6721
\item Greiner, J., Schwarz, R., Englhauser, J., Groot, P. J., 
        \& Galama, T. J. 1997b, IAU Circ. 6757
\item Heise, J. et al. 1997, IAU Circ. 6787
\item Huang, Y. F., Dai, Z. G., Wei, D. M., \& Lu, T. 1998, MNRAS, in press
\item Katz, J. 1994, ApJ, 422, 248
\item Katz, J., \& Piran, T. 1997, ApJ, 490, 772
\item Lang, K. R. 1980, Astrophysical Formulae (2d corr. and enl. ed.; 
        Berlin: Springer), 302
\item Marshall, F. E., Takeshima, T., Barthelmy, S. D., Robinson, C. R., 
        \& Hurley, K. 1997a, IAU Circ. 6683
\item Marshall, F. E., Gannizzo, J. K., \& Corbet, R. H. D. 1997b,
        IAU Circ. 6727
\item M\'esz\'aros, P., \& Rees, M. J. 1993, ApJ, 405, 278
\item M\'esz\'aros, P., \& Rees, M. J. 1997, ApJ, 476, 232
\item Murakami, T., Fujimoto, R., Ueda, Y., \& Shibata, R. 1997a, IAU Circ. 6687
\item Murakami, T., Ueda, Y., Ishida, M., \& Fujimoto, R. 1997b, IAU Circ. 6722
\item Murakami, T., Ueda, Y., Yoshida, A., Kawai, N., Marshall, F. E., 
        Corbet, R. H. D., \& Takeshima, T. 1997c, IAU Circ. 6732
\item Paczy\'nski, B., \& Rhoads, J. 1993, ApJ, 418, L5
\item Paczy\'nski, B., \& Xu, G. 1994, ApJ, 427, 708
\item Panaitescu, A., \& M\'esz\'aros, P. 1998, ApJ, 493, L31
\item Panaitescu, A., M\'esz\'aros, P., \& Rees, M. J. 1998, preprint:
        astro-ph/9801258
\item Pedersen, H. et al. 1998, ApJ, 496, 311
\item Piro, L. et al. 1997a, IAU Circ. 6617 
\item Piro, L. et al. 1997b, IAU Circ. 6656 
\item Piro, L. et al. 1997c, IAU Circ. 6797 
\item Piro, L. et al. 1998, A\&A, 331, L41
\item Rees, M. J., \& M\'esz\'aros, P. 1992, MNRAS, 258, 41p
\item Rees, M. J., \& M\'esz\'aros, P. 1994, ApJ, 430, L93
\item Rees, M. J., \& M\'esz\'aros, P. 1998, ApJ, 496, L1
\item Reichart, D. E. 1997, ApJ, 485, L57  
\item Remillard, R., Wood, A., Smith, D. A., \& Levine, A. 1997, IAU Circ. 6726
\item Rybicki, G. B., \& Lightman, A. P. 1979, Radiative Processes in
        Astrophysics, Wiley, New York
\item Sari, R. 1998, ApJ, 494, L49
\item Sari, R., \& Piran, T. 1997, ApJ, 495, 270
\item Sari, R., Piran, T., \& Narayan, R. 1998, ApJ, 497, L17
\item Shu, F. H. 1992, The Physics of Astrophysics, Vol. 2, Gas Dynamics 
        (Mill Valley, CA: Univ. Sci.)
\item Smith, D. A., Levine, A. M., Morgan, E. H., \& Wood, A. 1997, IAU Circ. 6718
\item Tavani, M. 1997, ApJ, 483, L87
\item Vietri, M. 1997a, ApJ, 478, L9
\item Vietri, M. 1997b, ApJ, 488, L105
\item Waxman, E. 1997a, ApJ, 485, L5
\item Waxman, E. 1997b, ApJ, 489, L33
\item Wijers, R. A. M. J., Rees, M. J., \& M\'esz\'aros, P. 1997,
        MNRAS, 288, L51
\item Yoshida, A. et al. 1997, IAU Circ. 6593
\end{description}

\newpage
\baselineskip=5.5mm

\begin{flushleft}
{\bf Table 1.} X-ray observations of recently localized GRBs. \\
\vspace{10mm}
\begin{tabular}{cccccc} \hline 
  GRB & Time Delay & Instrument & Energy Range & Flux & Ref.$^{(a)}$  \\
      &  log t(s)  &            &   (keV)      & (ergs cm$^{-2}$ s$^{-1}$)
      &  \\
\hline
970111 &  initial burst  &  BeppoSAX/WFC  &   2 $-$ 26   &  4 Crab  &  1 \\
\hline
970228 &  initial burst  &  BeppoSAX/WFC  &   2 $-$ 26   &  4 Crab  &  2 \\
  & 4.46 & BeppoSAX/MECS & 2 $-$ 10 & $(2.8 \pm 0.4) \times 10^{-12}$ & 3 \\
  & 4.46 & BeppoSAX/LECS & 0.5 $-$ 10 & $(4.0 \pm 0.6) \times 10^{-12}$ & 3 \\
  & 5.49 & BeppoSAX/NFI  & 2 $-$ 10 & $1.4 \times 10^{-13}$ & 3 \\
  & 5.79 & ASCA/GIS      & 2 $-$ 10 & $(9.0 \pm 2.6) \times 10^{-14}$ & 4 \\
  & 5.79 & ASCA/SIS      & 2 $-$ 10 & $(7.2 \pm 2.1) \times 10^{-14}$ & 4 \\
  & 6.02 & ROSAT/HRI  & 0.1 $-$ 2.4 & $(3.8 \pm 1.2) \times 10^{-14}$ & 5 \\
\hline
970402 &  initial burst  &  BeppoSAX/WFC  &   2 $-$ 26   & 0.46 Crab  & 6 \\
  & 4.48 & BeppoSAX/MECS & 2 $-$ 10 & $(1.5 \pm 0.5) \times 10^{-13}$ & 7 \\
  & 4.48 & BeppoSAX/LECS & 0.5 $-$ 5 & $(2.0 \pm 0.6) \times 10^{-13}$ & 7 \\
  & 5.17 & BeppoSAX      & 2 $-$ 10  & $< 5 \times 10^{-14}$           & 7 \\
\hline
970508 &  initial burst  &  BeppoSAX/WFC  &   2 $-$ 26   & 1 Crab  & 8 \\
  & 4.30 & BeppoSAX/MECS & 2 $-$ 10 & $(6.3 \pm 0.6) \times 10^{-13}$  & 9 \\
  & 4.30 & BeppoSAX/LECS & 0.5 $-$ 5 & $(7.0 \pm 0.8) \times 10^{-13}$ & 9 \\
  & 4.38 & BeppoSAX/MECS & 2 $-$ 10 & $ 4 \times 10^{-13}$    & 10 \\
  & 5.55 & BeppoSAX/MECS & 2 $-$ 10 & $ 2.4 \times 10^{-13}$  & 10 \\
\hline 
970616 & 4.16  & RXTE & 2 $-$ 10 & $1.1 \times 10^{-11}$ & 11 \\ 
       & 5.54  & ASCA & 2 $-$ 7  & $6.9 \times 10^{-14}$ & 12 \\ 
     & 5.80  & ROSAT & 0.5 $-$ 2 & $1.4 \times 10^{-14}$ & 13 \\ 
\hline
970815 & initial burst  & RXTE/ASM & 2 $-$ 10 & 2 Crab & 14 \\ 
       & 5.51  & ASCA & 2 $-$ 10  & $ < 1 \times 10^{-13}$ & 15 \\ 
     & 5.68  & ROSAT & 0.1 $-$ 2.4 & $5 \times 10^{-14}$ & 16 \\ 
\hline 
970828 & 1.30  & RXTE & 2 $-$ 12 & 0.756 Crab & 17 \\ 
       & 1.95  & RXTE & 2 $-$ 12 & 0.238 Crab & 17 \\ 
       & 4.11  & RXTE & 2 $-$ 10 & $ 5 \times 10^{-4}$ Crab & 18 \\ 
       & 5.15  & ASCA & 2 $-$ 10  & $ 4 \times 10^{-13}$ & 19 \\ 
     & 5.80  & ROSAT/HRI & 0.1 $-$ 2.4 & $2.5 \times 10^{-14}$ & 20 \\ 
\hline
971214 &  initial burst  &  BeppoSAX/WFC  &   2 $-$ 26   & 1 Crab  & 21 \\
  & 4.38 & BeppoSAX/MECS & 2 $-$ 10 & $ 4 \times 10^{-13}$  & 22 \\
\hline 
971227 &  initial burst  &  BeppoSAX/WFC  &   2 $-$ 26   & 1.8 Crab  & 23 \\
  & 4.81 & BeppoSAX/MECS & 2 $-$ 10 & $ 3 \times 10^{-13}$  & 24 \\
\hline
\end{tabular}

\noindent 
$^{(a)}$ (1) Costa et al. 1997a; (2) Costa et al. 1997b; (3) Costa et al. 
1997c; (4) Yoshida et al. 1997; (5) Frontera et al. 1997; (6) Feroci et al. 
1997; (7) Piro et al. 1997a; (8) Costa et al. 1997d; (9) Piro et al. 1997b; 
(10) Greiner 1998; (11) Marshall et al. 1997a; (12) Murakami et al. 1997a; 
(13) Greiner et al. 1997a; (14) Smith et al. 1997; (15) Murakami et al. 
1997b; (16) Greiner 1997; (17) Remillard et al. 1997; (18) Marshall et al. 
1997b; (19) Murakami et al. 1997c; (20) Greiner et al. 1997b; (21) Heise 
et al. 1997; (22) Antonelli et al. 1997; (23) Coletta et al. 1997; 
(24) Piro et al. 1997c.

\end{flushleft}

\newpage
\baselineskip=6mm

\begin{center}
FIGURE CAPTIONS
\end{center}

\noindent
{\bf Figure 1.} Evolution of the fireball's kinetic energy. The solid 
lines is drawn with ``standard'' parameters ($E  = 10^{51}$ ergs, 
$n = 1$ cm$^{-3}$, $M_0 = 2 \times 10^{-6}$ $M_{\odot}$, 
$\xi_B^2 = 0.01$, $\xi_e = 1.0$). Each of the other lines is plotted 
with one parameter altered. The dotted line corresponds to 
$M_0 = 1 \times 10^{-6}$ $M_{\odot}$, the dashed line corresponds 
to $\xi_B^2 = 0.1$, and the dash-dotted line corresponds to
$\xi_e = 0.5$.

\noindent
{\bf Figure 2.} Evolution of the fireball's Lorentz factor. Parameters
are the same as in Figure 1.

\noindent
{\bf Figure 3.} Evolution of the fireball's radius. Parameters are the
same as in Figure 1.

\noindent
{\bf Figure 4.} Velocity of the blastwave ($\log V/c$) vs. the fireball's
radius. Full line is our numerical result, plotted with the 
``standard'' parameters as in Figure 1. Dotted line is the analytic 
result for non-relativistic phase, 
$R^3 = 1.15^5 \times 4 E /(25 n V^2)$ (Lang 1980; Shu 1992), 
in which we have set $E = 7.2 \times 10^{49}$ ergs, since this is 
approximately the kinetic energy left in a non-relativistic
fireball for our model (see Figure 1).

\noindent
{\bf Figure 5.} Evolution of the radiative efficiency of the fireball. 
Parameters are the same as in Figure 1.

\noindent
{\bf Figure 6.} Evolution of the index $\alpha$, where $\alpha$ is 
defined as $\gamma \propto R^{- \alpha}$. Parameters are the same as 
in Figure 1.

\noindent
{\bf Figure 7.} Evolution of the mass of the swept-up medium. Parameters
are the same as in Figure 1.

\noindent
{\bf Figure 8.} Predicted optical afterglows in R band, $S_R$. We take 
$p = 2.1$, $D = 3$ Gpc (a) or $D = 1.5$ Gpc (b). Other parameters are 
the same as in Figure 1. Also plotted are the observed afterglows from 
GRB 970228 (a) (Galama et al. 1998a; Fruchter et al. 1997; Garcia et al. 
1997; Wijers, Rees, \& M\'{e}sz\'{a}ros 1997), from GRB 970508 (b) 
(Pedersen et al. 1998; Garcia et al. 1997; Galama et al. 1998a).

\noindent
{\bf Figure 9.} Predicted X-ray afterglows (0.1--10 keV). 
The flux is in unit of  ergs cm$^{-2}$ s$^{-1}$. We take
$D = 3$ Gpc. Other parameters are the same as in Figure 8.
Also plotted are the observed 
data, which have been linearly scaled to $0.1 - 10$ keV fluxes from 
Table 1.

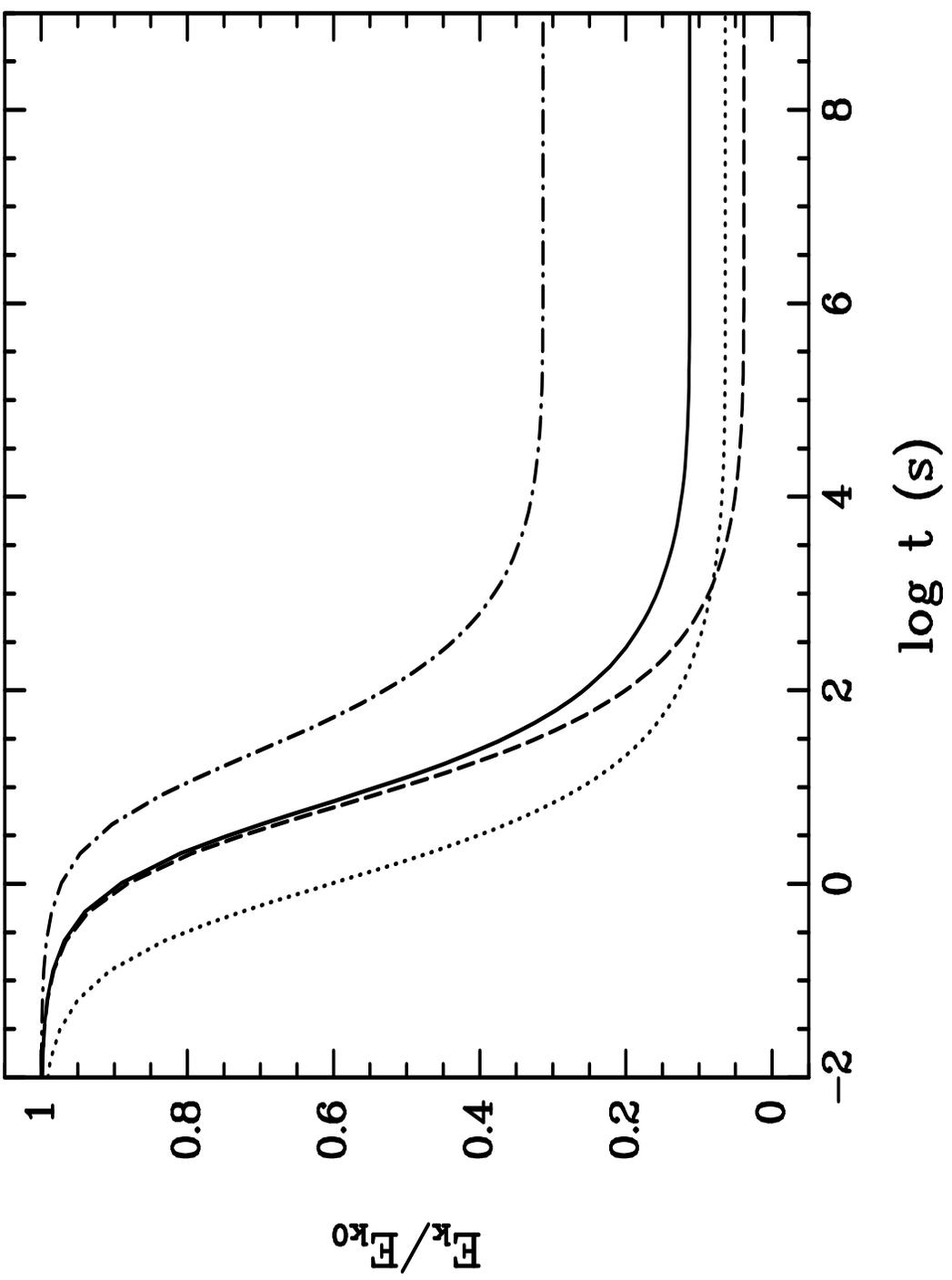
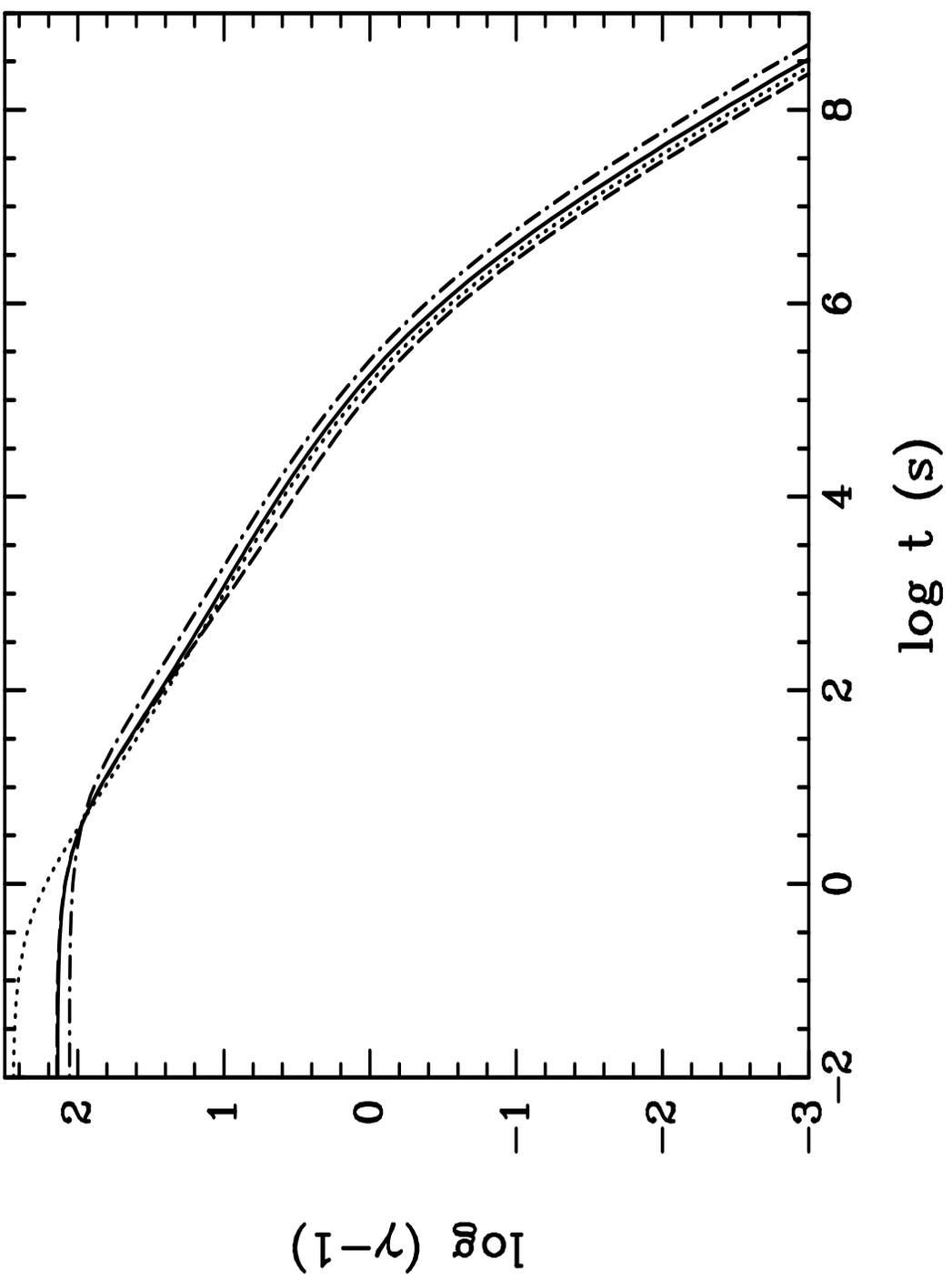
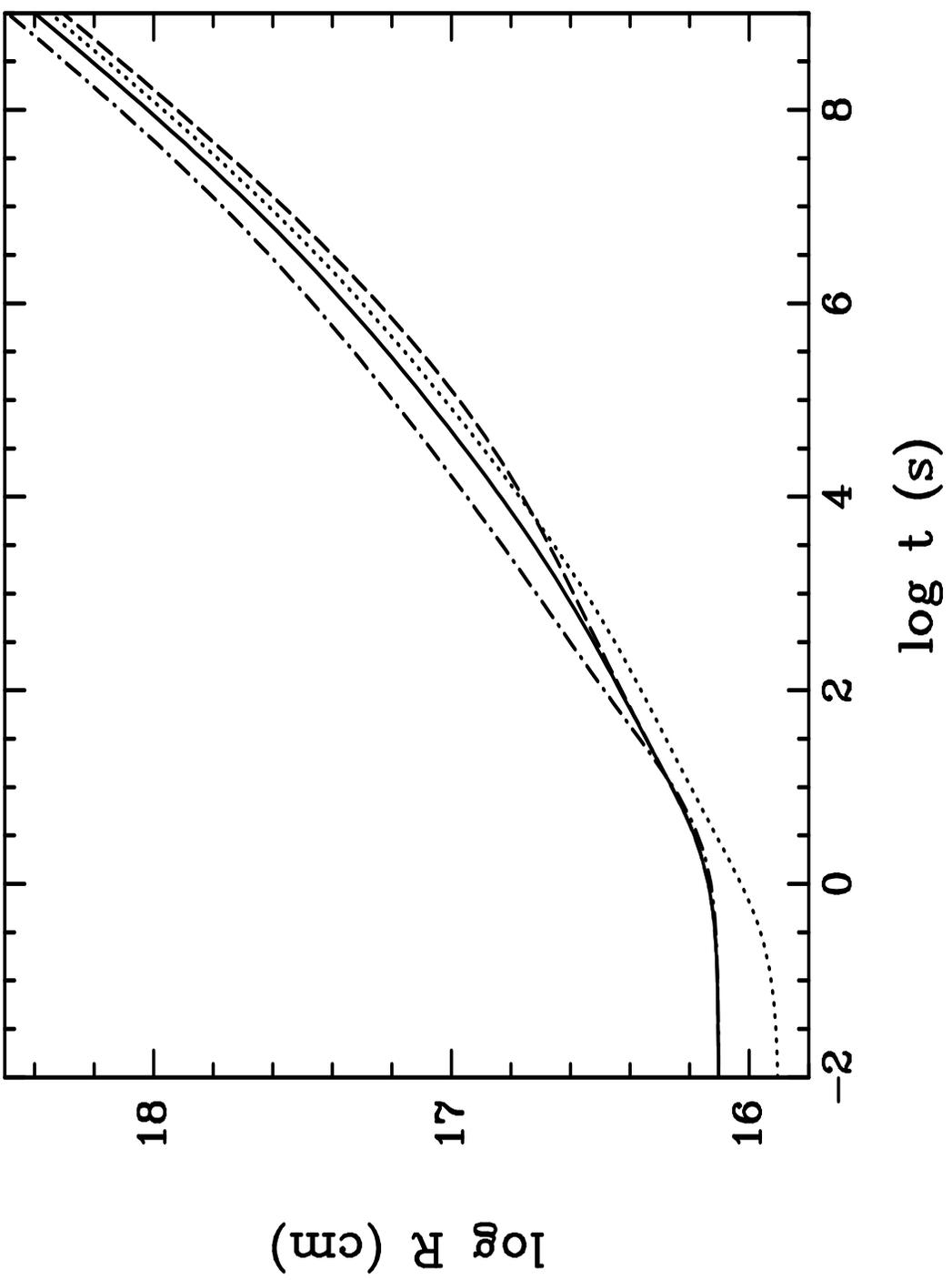
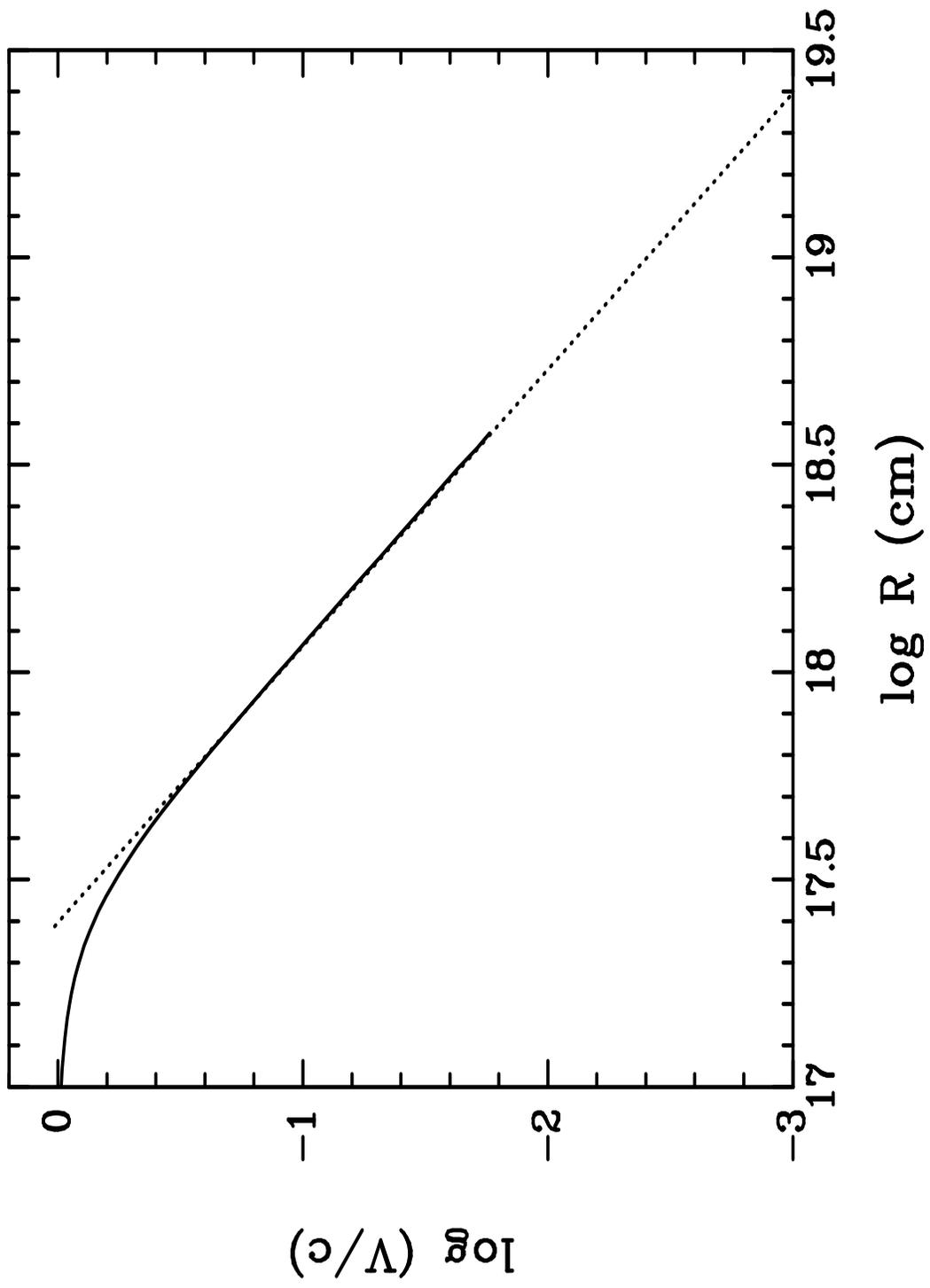
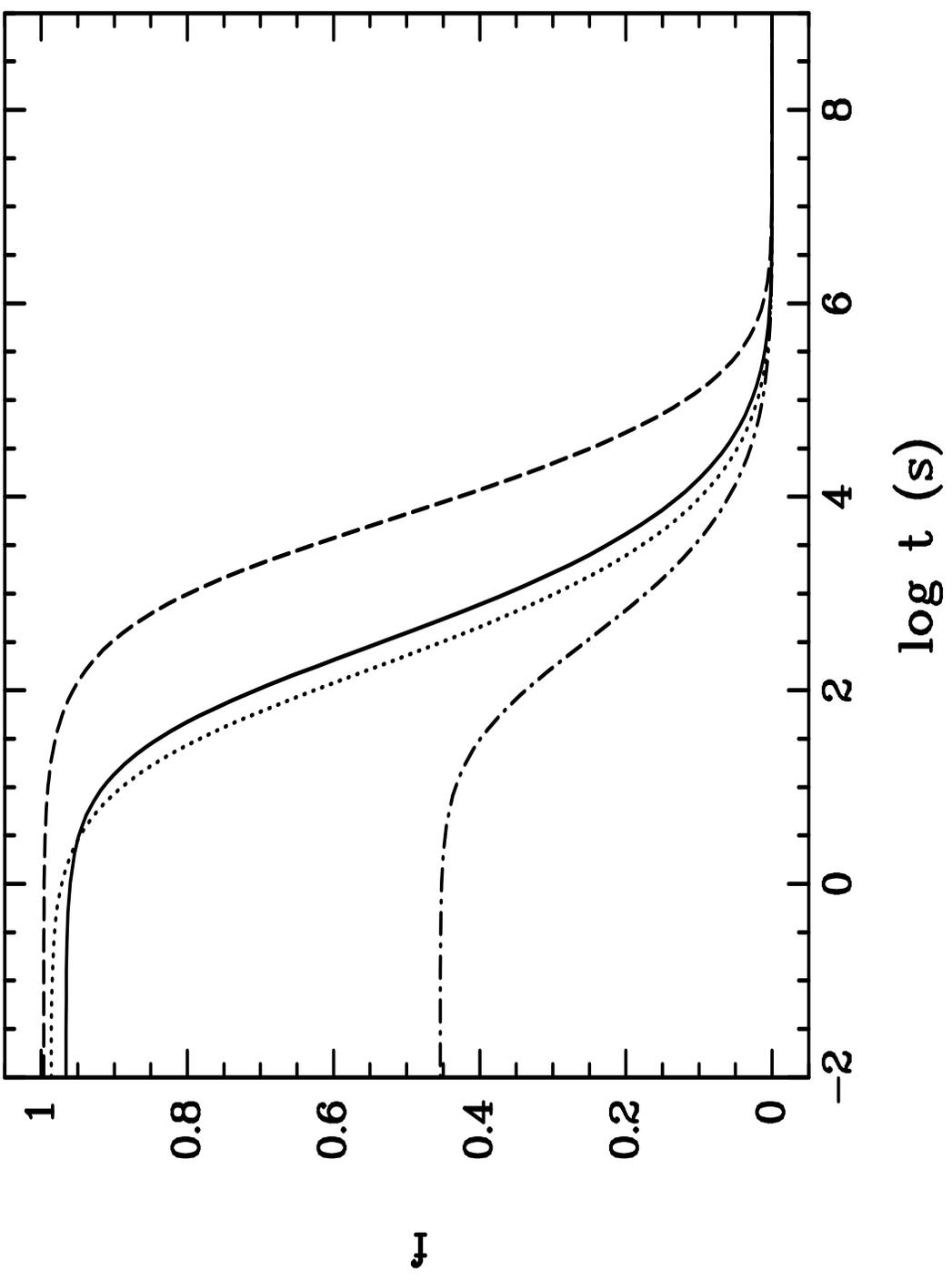
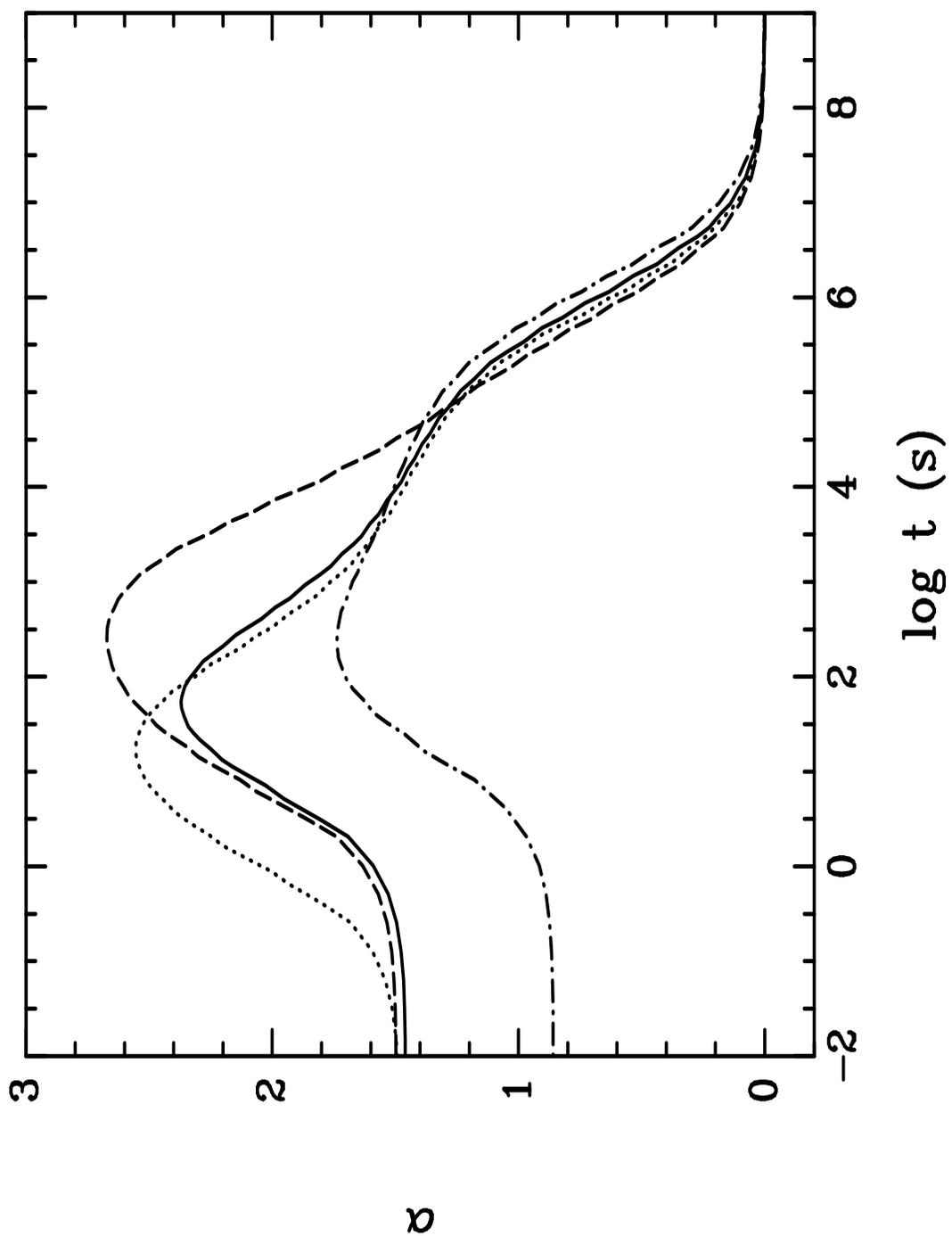
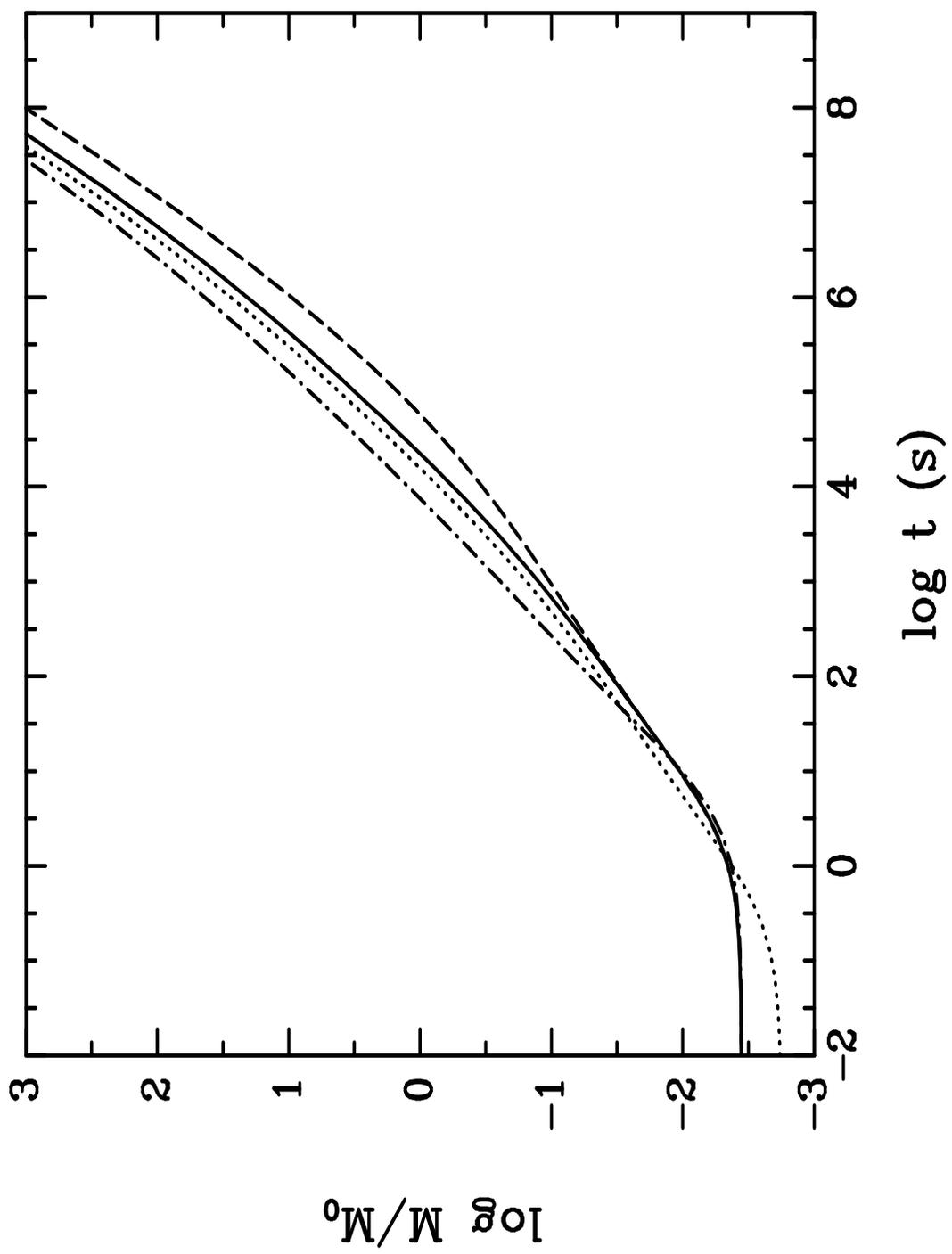
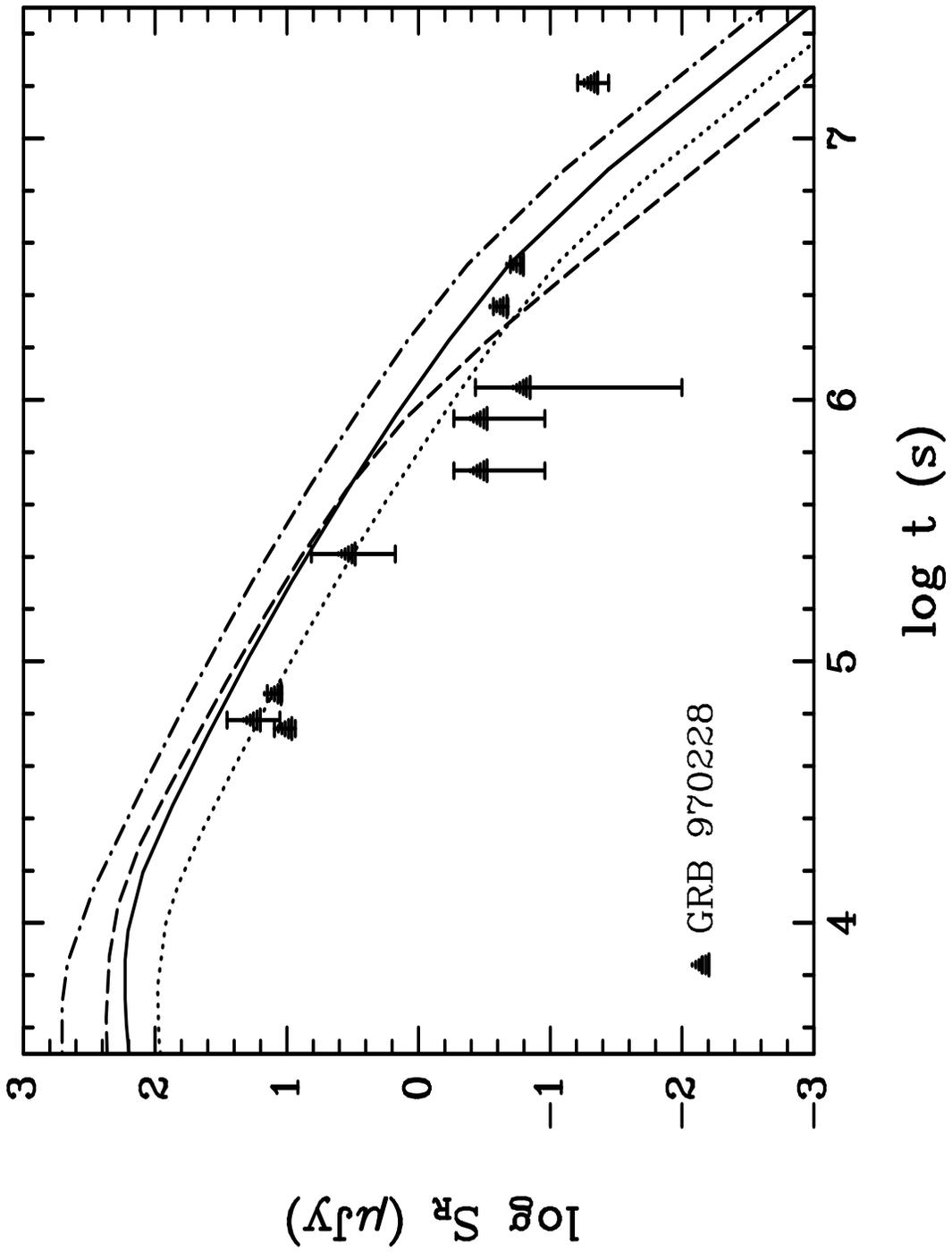
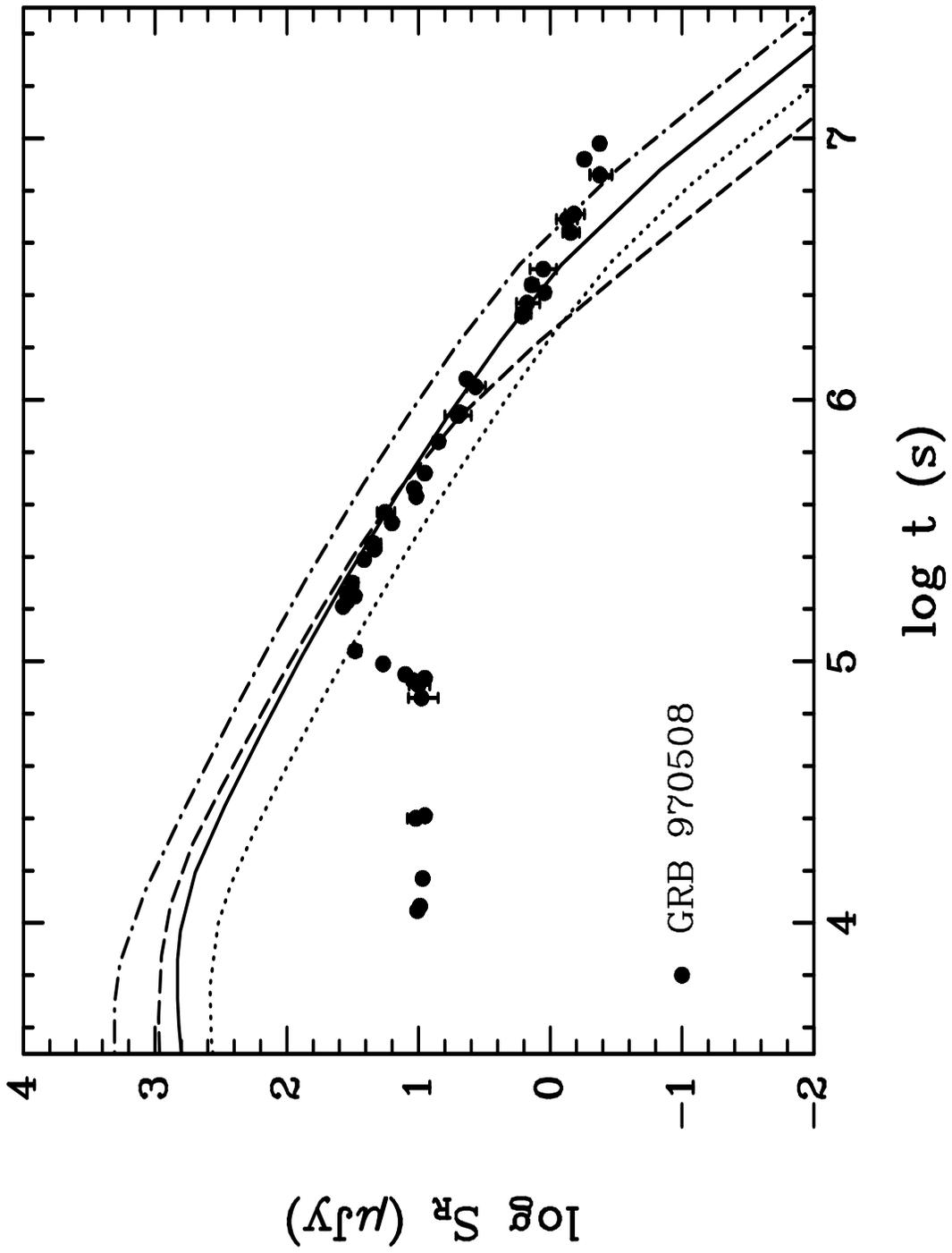
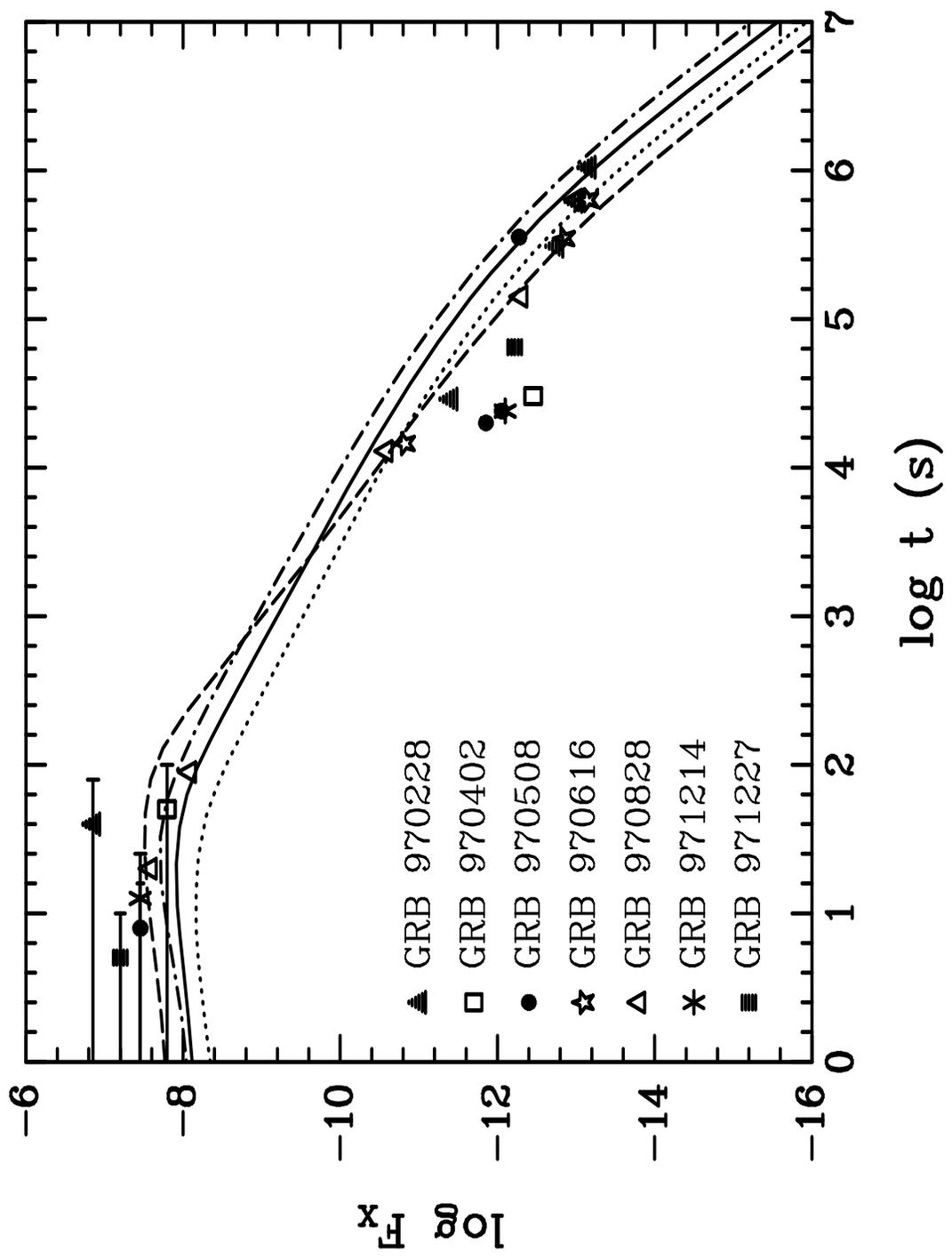
\end{document}